\documentstyle[11pt,fleqn,epsfig]{article}

\begin{document}

 \title{
 The magnetic reversal 
 in dot arrays 
 recognized by 
 the self-organized 
 adaptive neural network
 }
 \author{     Martin Gmitra and Denis Horv\'ath
              \\
              Department of Theoretical Physics and Astrophysics, \\
              University of P.J.\v{S}af\'arik,                    \\
              Moyzesova 16, 040 01 Ko\v{s}ice,                    \\
              Slovak Republic                                     \\
       }

 \date{}
 \maketitle

 \begin{abstract}
The remagnetization dynamics of monolayer 
dot array superlattice XY 2-D spin model 
with dipole-dipole interactions is simulated. 
Within the proposed model of array, the square dots 
are described by the spatially modulated 
exchange-couplings. 
The dipole-dipole interactions 
are approximated by the hierarchical sums and spin dynamics 
is considered in  regime of the Landau-Lifshitz equation.
The simulation of reversal for $40 000$ spins 
exhibits 
formation of nonuniform intra-dot configurations 
with nonlinear wave/anti-wave pairs developed 
at intra-dot and inter-dot scales. 
Several geometric and parametric dependences 
are calculated and compared with oversimplified four-spin 
model of reversal. The role of initial conditions 
and the occurrence of coherent 
rotation mode is also investigated. 
The emphasis is on the classification 
of intra-dot or inter-dot (interfacial) 
magnetic configurations done by adaptive neural 
network with varying number of neurons.
\end{abstract}

 \vspace{0.3cm}

 \noindent PACS:  
 Nanostructures: 79.60.J,  
 Magnetostatics 41.20.G,
 Magnetic ordering 75.10,  
 Numerical simulation studies 75.40.Mg

 \vspace{0.3cm}


 \vspace{0.2cm}

\section{Introduction}

Many interesting applications of the nanometer periodic magnetic
dot arrays are expected in the magnetic recording
\cite{Prinz95} and magnetic sensors. 
The magnetic nanostructures represent model systems for study 
of interesting physical phenomena stemming from the well controlled 
dot shape anisotropy and dominance of magnetostatic interactions 
at the scales comparable or larger than dot size.
The magnetostatics cause the coupling of domains over 
the inter-dot interfaces. 
The current perspective technologies
require understanding and controlling of dynamic magnetization 
processes in fine particles on nanoscales, 
because by them the magnetic 
bits are written into recording media. 
Hence of particular interest
is the dynamics of magnetization reversal.

The static and dynamic 
nonuniform magnetization states
in magnetic nanostructures
including the thin films and isolated small particles 
have been studied 
by micromagnetic {\cite{Natali2002,Liou2001,Lieneweber1999}}
and 
Monte Carlo
\cite{Levy1999,Iglesias2000,Kachkachi2000,Santamaria2002}
simulations. On a qualitative level intra-dot 
nonuniformities 
can be characterized as a superpositions 
of vortices, flowers, domain walls, etc..
The problem of nonuniformities in exchange-dominated  
structures can be solved exactly \cite{Metlov2002}.
 
Our recent activity is directed to algorithms 
allowing systematic and fast classification of simulated 
magnetic structures.
The methodology presented in this paper
is inspired by the progress in the theory of artificial
neural networks that are nonlinear models suitable to
reduce, classify or interpolate data structures 
representing the complex patterns {\cite{Haykin98}}.
In the previous paper {\cite{Horvath2001}}
we suggested the implementation 
of radial basis function
networks to model the ordering of dots in arrays.  
In the next paper \cite{Horvath2002} 
the neural network 
training was applied to 
multiscale numerical computations. 
The lack of our previous two approaches 
was the absence of realistic 
dynamics and restriction to quasi-static limit. In the present 
paper our interest is focused to the networks which 
analyze spin configurations generated 
by the spin lattice dynamical simulations. 
It means that network solves here classical 
post-processing problem 
known in the micromagnetic context 
\cite{Schabes87}. Our approach 
has many in common with 
experiment treatment \cite{Jones2002},
where self-organized network 
was employed to identification of vortex 
modes of magnetic thin film media.  
In the presented paper the classification 
is applied to configurations generated 
by the magnetization reversal 
within the XY superlattice model 
of ultra-dense array. 
The motivation for study 
of ultra-dense spin 
structures was the expectation 
of extremal 
inter-dot effects. 

The main reason for the choice 
of the XY spin model is relative 
computational simplicity 
of 2-D dynamics. 
But there are two regimes for 
which 2-D approximation 
is justified physically.
(i)~The first regime belongs 
to limit of very strong damping
which suppresses precession 
and accelerates reversal \cite{Kikuchi}.
(ii)~The second regime 
represents limit of strong in-plain \cite{Kalmykov2000} 
or shape anisotropy associated with  demagnetization field of thin 
film sample \cite{Braun93}. In this regime anisotropy 
causes deformation of spin trajectory and accelerates 
reversal \cite{Bauer2000}. Nevertheless, one should 
to note that realistic damping is sufficiently small, 
and thus 2-D remagnetization 
is practically quasistatic. In the quasistatic 
limit the damping parameter has meaning of auxiliary 
parameter rescaling the time. 
Recent results 
obtained for discrete spin systems 
provide more optimistic view 
on the valuation of artificial 
dynamics of reversal. 
The example is the simulation of \cite{Hinzke221} 
demonstrating that switching 
time Monte Carlo dependences 
can coincide after 
some rescaling with predictions of the 
stochastic Landau-Lifshitz equation.

The plan of the paper is the following.
The interaction picture and dynamics 
of array model are defined in 
section~\ref{xy_model}. 
The simulations of array reversal 
are presented and discussed 
in section~\ref{results1}. 
The details of adaptive 
neural network treatment 
of intra-dot configurations 
are given in subsection~\ref{ART1}.

\section{XY periodically modulated superlattice model 
of array}\label{xy_model}

In this section we formulate the interaction picture
and dynamics of 
ultra dense array model.
The magnetic ordering is described by the  
systems of the classical planar
spins located on the square 
two dimensional lattice.
The spin on site $\alpha$ 
is affected by effective
field
\begin{equation}
{\bf h}_{\alpha}^{\rm eff} 
=
\frac{D}{J}
\sum_{{\beta=0}
\atop{ \beta
\not=\alpha}}^{L^2-1}
              \frac{
3 {\bf n}_{\alpha\beta}
       ({\bf n}_{\alpha\beta}\cdot{\bf S}_\beta)
                     - {\bf S}_\beta
                   }
                   { |{\bf r}_{\alpha\beta}|^3 }
              +
\sum_{{\beta=0}
\atop
{|{\bf r}_{\alpha\beta}|=1}}^{L^2-1}
\epsilon_{\alpha\beta} 
{\bf S}_{\beta} + \frac{\bf H}{J}\,,
\label{ExchEq}
\end{equation}
where ${\bf S}_{\alpha}=
S_{\alpha{\rm x}}{\bf e}_{\rm x}+ 
S_{\alpha{\rm y}} {\bf e}_{\rm y}$
is the planar spin,
${\bf H}$ is the external magnetic field in energy units,
$L$ is the linear size of the square 
$L\times L$ spin lattice system.
Here and in further  
${\bf e}_{\rm x}$, 
${\bf e}_{\rm y}$
denote the unit vectors along the main Cartesian axes.
Rather convenient for a computer purposes is to use 
one dimensional enumeration scheme of spin sites
$\alpha\in \{0,1,2,\ldots L^2-1\}$. 
(All of the geometric details presented in this section
are discussed in the Appendix~\ref{1d_enumeration}.)
The first sum of Eq.(\ref{ExchEq}) 
represents dipolar contribution proportional 
to constant $D$, 
where ${\bf r}_{\alpha\beta}$ 
is vector pointing from site 
$\alpha$ to $\beta$. The vector is measured 
in lattice constant units, 
${\bf n}_{\alpha\beta}={\bf r}_{\alpha\beta}/|{\bf r}_{\alpha\beta}|$ 
is the auxiliary unit vector. 
The second sum 
of Eq.(\ref{ExchEq}) involves nearest 
neighbor exchange coupling term
$J \epsilon_{\alpha\beta}$ 
determined by single parameter 
$J>0$ multiplying the site-site connectivity matrix 
$\epsilon_{\alpha\beta} \in \{1,0\}$
constrained by 
$|{\bf r}_{\alpha\beta}|=1$ condition; 
$\epsilon_{\alpha\beta}$
is $0$ for any bond at inter-dot boundary 
and $1$ inside a dot. 
The site dependence 
of $\epsilon_{\alpha\beta}$ 
takes into account square shape 
of dot of the linear size $L_0$.
The spin system partitioned by 
$\epsilon_{\alpha\beta}$ consists of 
$L_{\rm d}^2=(L/L_0)^2$ dots with minimum 
inter-dot distance equal to the unit lattice spacing 
and corresponding to extremal space filling. 

In 2-D case Landau-Lifshitz equation reduces to 
damping term. The numerical integration 
scheme discussed in Appendix \ref{integ_scheme} 
can be derived from recursive 
propagator form
\begin{equation}
\label{intscheme}
{\bf S}_\alpha(\tau+\Delta \tau) =
\hat{\bf U}\left(
\int_\tau^{\tau+\Delta \tau}
{\omega}_\alpha(\tau')
\,
\mbox{d} \tau'
\,
\right)
\,{\bf S}_\alpha(\tau)\,,
\end{equation}
where ${\omega}_\alpha(\tau)
=(   {\bf S}_\alpha(\tau) \cdot {\bf e}_{\rm x}              )
 (   {\bf h}^{\rm eff}_\alpha(\tau)  \cdot {\bf e}_{\rm y}   ) -
 (   {\bf S}_\alpha (\tau) \cdot {\bf e}_{\rm y}             )
 (   {\bf h}^{\rm eff}_\alpha(\tau)  \cdot {\bf e}_{\rm x}   ) $
is instant angular frequency of the spin rotation,
$\Delta \tau$ is the time integration step, 
$\tau$ is the reduced time related to real time $t$ as 
$\tau=\mu_0\gamma M_{\rm s} t$, where 
$\mu_0$ is permeability of free space, $\gamma$ 
is the gyromagnetic ratio 
and $M_{\rm s}$ is the saturation magnetization.
The unitarity of spin rotation matrix
\begin{equation}
\label{rot_matrix}
 \hat{\bf U}(\varphi) = \left(
                     \begin{array}{lr}
                      \cos\varphi & -\sin\varphi \\
                      \sin\varphi & \cos\varphi
                     \end{array}
                     \right)
\end{equation}
implies that numerical scheme adapted
from Eq.(\ref{intscheme}) 
conserves the spin vector 
size independently of 
$\int_\tau^{\tau+\Delta \tau} {\omega}_\alpha(\tau') \mbox{d} \tau'$
quadrature.

The time demanding task represents summation of dipolar fields.
The substantial reduction of time allows 
the use of the hierarchical approach 
\cite{Miles91} with a detailed summation 
of near and rough summation of far field 
contributions. The specific geometry 
of array leads to 
the natural spin block 
definition
\begin{equation}
{\bf S}^{\rm d}_{j}=\sum_{\beta=0\atop i(\beta)=j}^{L^2-1}
{\bf S}_{\beta}\,,\qquad j\in \{0,1,2,\ldots,L_{\rm d}^2-1\}\,,
\end{equation}
where sum runs over the spins of index $\beta$ belonging 
to dot index $i(\beta)=j$ and dot position
$\,{\bf r}^{\rm d}_{i(\beta)}$.
In frame of
two-level hierarchical approximation
dipolar field is splitted to near
${\bf h}^{\rm eff,N}_{\alpha}$
and far
${\bf h}^{\rm eff,F}_{\alpha}$
field
\begin{equation}
\label{h_near}
{\bf h}^{\rm eff,N}_{\alpha} = 
\frac{D}{J}
\sum_{
{\beta=0\, ,\, \beta\not=\alpha}
\atop {|{\bf r}^{\rm d}_{i(\alpha),i(\beta)}|\leq r_{\rm cut}}}^{L^2-1}
       \frac{
                     3 \, {\bf n}_{\alpha\beta}
                      ({\bf n}_{\alpha\beta}\cdot {\bf S}_\beta)
                     - {\bf S}_\beta
                   }
                   { |{\bf r}_{\alpha \beta}|^3 }
+
\sum_{
{\beta=0}
\atop
{|{\bf r}_{\alpha \beta}|=1}
}^{L^2-1}
{\epsilon}_{\alpha\beta}{\bf S}_\beta
             +\frac{\bf H}{J}\,,
\end{equation}

\begin{equation}
\label{h_far}
{\bf h}^{\rm eff,F}_{\alpha} = \frac{D}{J}
\sum_{
{j=0}
\atop {
|{\bf r}^{\rm d}_{i(\alpha)j}|>r_{\rm cut}}
}^{L^2_{\rm d}-1}
\,\,\,
\frac{
3\,{\bf n}^{\rm d}_{i(\alpha)\, j}\,
({\bf n}^{\rm d}_{i(\alpha)\, j}
\cdot {\bf S}^{\rm d}_{j})
- {\bf S}^{\rm d}_{j}
}
{|{\bf r}^{\rm d}_{i(\alpha)j}|^3 }\,\,,
\end{equation}
depending on the unit vectors
${\bf n}^{\rm d}_{i(\alpha)\,j}=
{\bf r}^{\rm d}_{i(\alpha)\,j}/
|{\bf r}^{\rm d}_{i(\alpha)\,j}|$,
${\bf r}^{\rm d}_{i(\alpha)\,j}=
{\bf r}^{\rm d}_{j}-{\bf r}^{\rm d}_{i(\alpha)}$.
The summation rule is controlled 
by the scale-separation cut-off 
$r_{\rm cut}$.

\section{The results of 
         array reversal 
         simulations}\label{results1}
In this section 
we present the simulations of the square shaped finite 
array including $40 000$ spins.
To analyze the simulation outputs we applied:  
(i)~the qualitative recognition of the configuration 
    snapshots;
(ii)~the comparing 
     of the complex array 
     dynamics with the dynamics 
     of much simpler systems;
(iii)~the clustering
      of intra-dot magnetic 
      configurations using neural networks.
The global information 
on the array ordering 
concerns the magnetization per site
\begin{equation}
\label{magnetization}
{\bf M}(\tau)=
\frac{1}{L^2}
\sum_{\alpha=0}^{L^2-1} 
{\bf S}_{\alpha}(\tau)\,.
\end{equation}
The simulations 
were performed for uniform zero
${\bf S}_{\alpha}|_{\tau=0}=-{\bf H}/H={\bf e}_{\rm x}$  
and non-zero bias  
${\bf S}_{\alpha}|_{\tau=0}=\cos(\varphi_0) 
{\bf e}_{\rm x}+\sin(\varphi_0) 
{\bf e}_{\rm y}$, $\varphi_0\neq 0$
initial conditions.  
In both cases the characteristic time 
of array is the time when ${\bf H} \cdot {\bf M}$ projection 
gains at first 
the bound  
\begin{equation}
{\bf H}\cdot{\bf M}(\tau_{\rm ch})=0\,.
\end{equation}

\subsection{The qualitative approach based on two-scale similarity}

To obtain 
the preliminary qualitative information about 
the switching mechanism 
we run the simulation of reversal dynamics 
in the field $H/J=-1$ and coupling $D/J=0.1$ 
for dot size $L_0=20$ and bias $\varphi_0=0$. 
Under these conditions we obtained
estimate $\tau_{\rm ch}=8.516$. 

\begin{figure}[h]\begin{center}
\epsfig{file=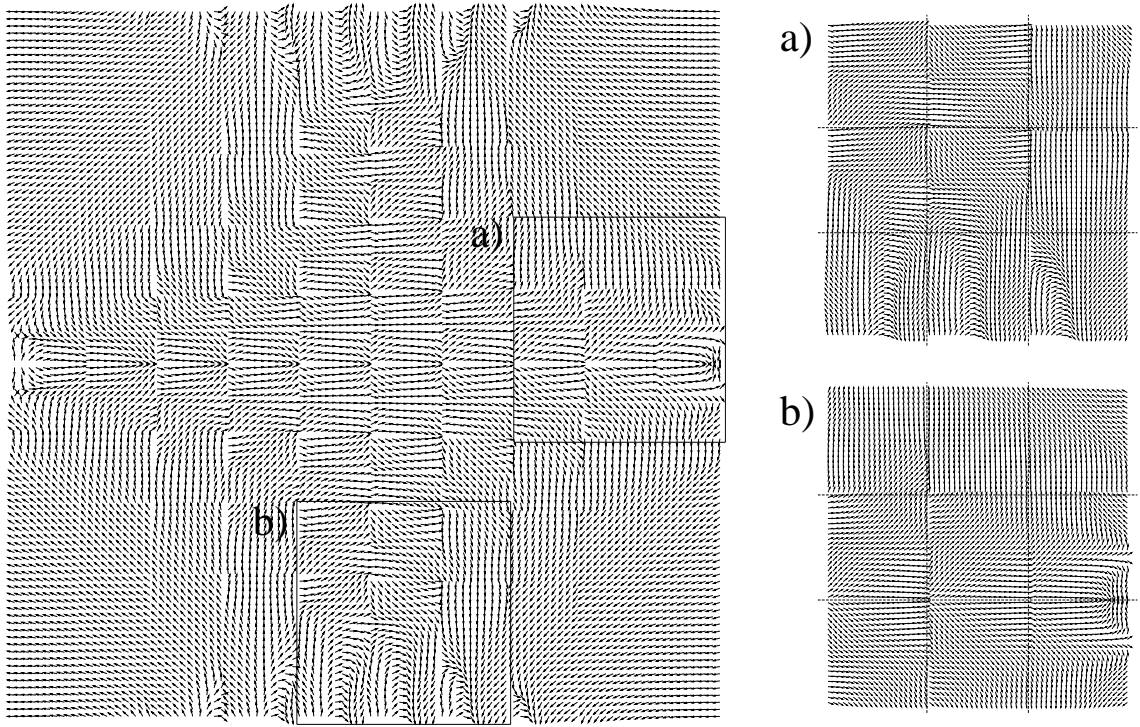,width=16.0cm,angle=0.0}
\caption{\label{cnf1}
The array snapshot including configuration with the resolution given 
by $N_{\rm S}=4=2\times 2$.  Obtained for parameters 
$D/J=0.1$, $L_0=20$ 
at characteristic time 
$\tau_{\rm ch}=8.516$. 
The differences between coherent 
rotation at array corners 
and flower mode 
near to the array center are striking. 
The dashed lines inside magnified segments~a), b) belong 
to bounds separating dots. 
Along these lines the exchange 
couplings are broken.
}
\end{center}\end{figure}

\begin{figure}[h]\begin{center}
\epsfig{file=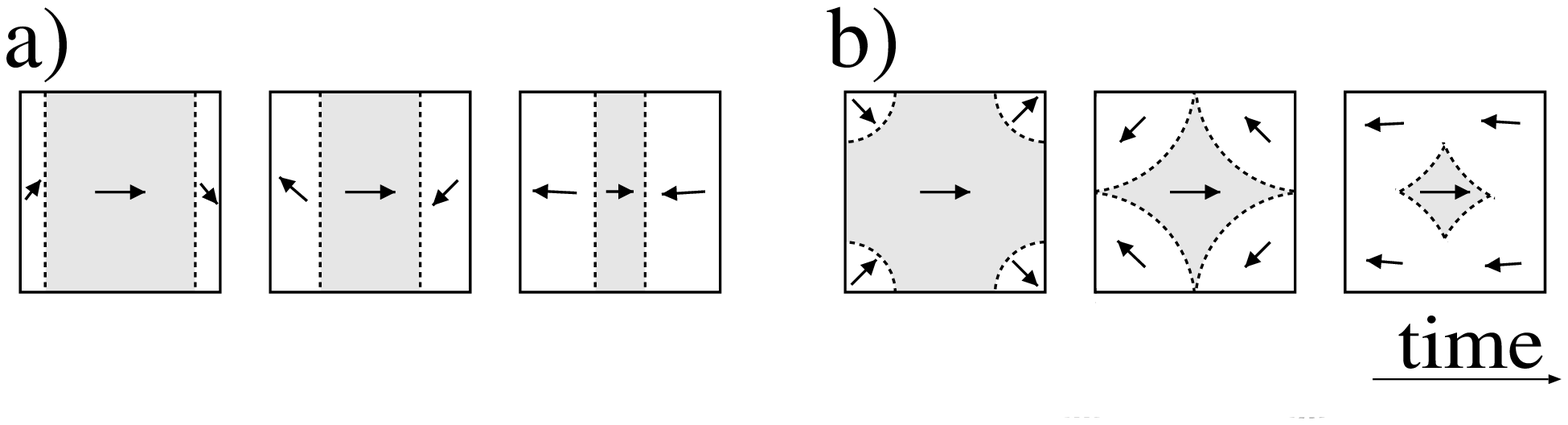,width=15.0cm,angle=0.0}
\caption{\label{model_premag}
Two schemes showing 
principal transients of nucleation, 
WAP formation and annihilation. 
The intra-dot open (straight) WAP (case~a) 
occurring 
at earlier transients 
of remagnetization in eccentric dots. 
The center-symmetric WAP closed mode 
(case~b) typical for late transients and dots 
positioned near the array center.}
\end{center}
\end{figure}

\begin{figure}[h]\begin{center}
\epsfig{file=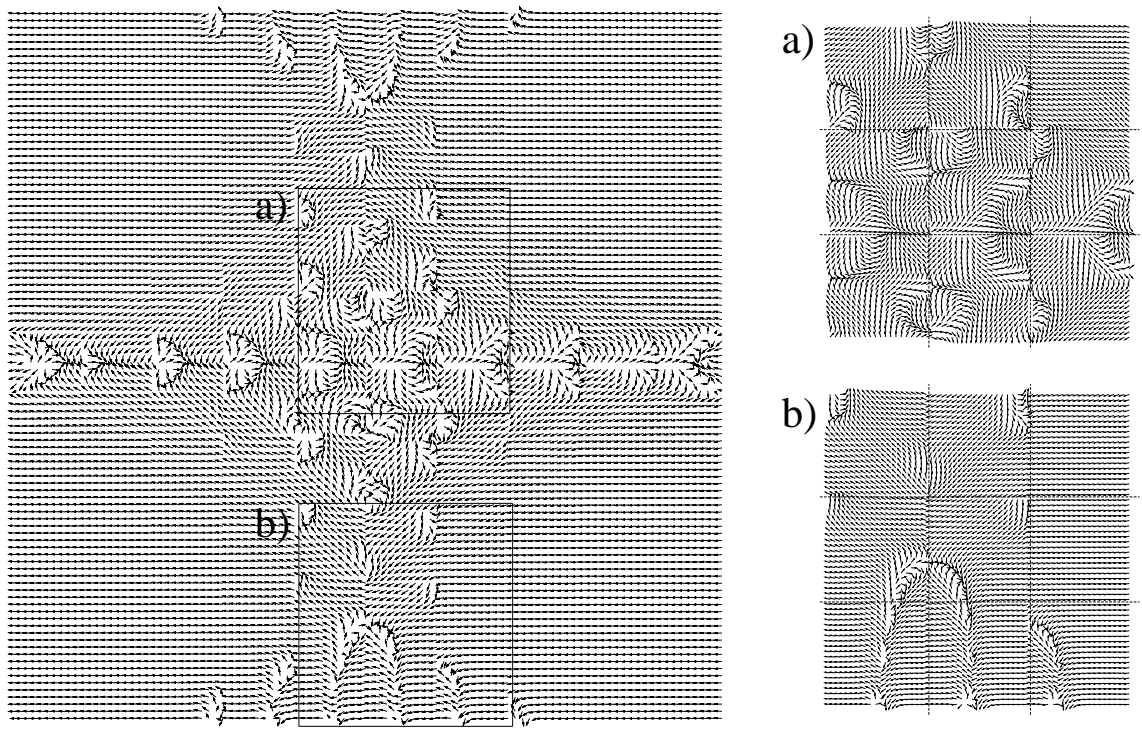,width=16.0cm,angle=0.0}
\caption{\label{cnf2}
 The array snapshot and its magnified details a), b)
 obtained for late stage ($\tau=11$)
 of reversal 
 (the dot labels and model parameters 
 are identical as in the Fig.~\ref{cnf1})
 nearly the time   
 when the internal energy 
 (free from Zeeman term) reaches 
 its maximum. 
 }\end{center}\end{figure}

\begin{figure}[h]\begin{center}
\epsfig{file=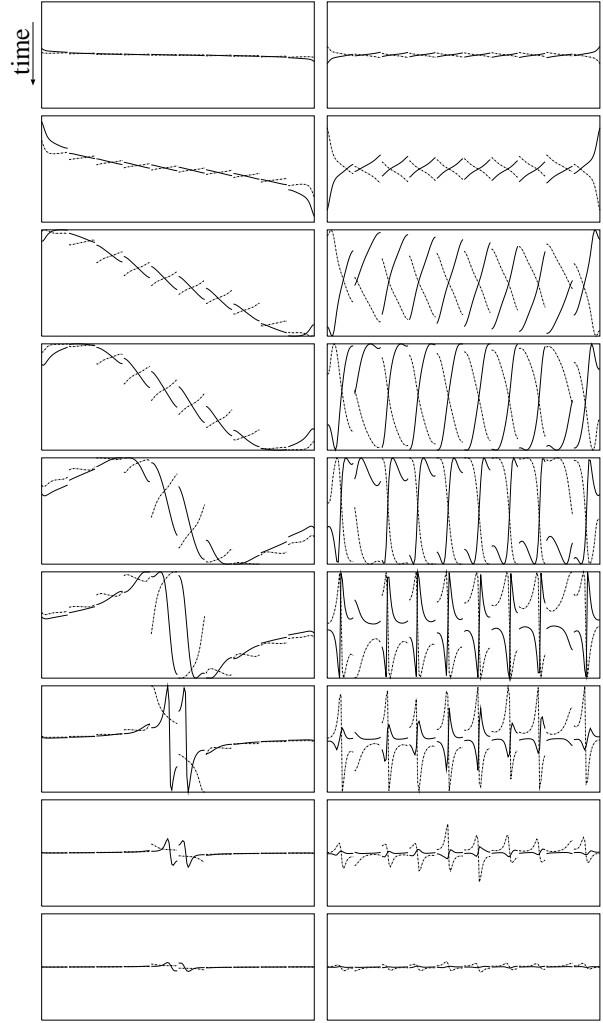,width=8.0cm,angle=0.0}
\caption{\label{solitons}
The reversal in the representation
of one dimensional spin projections 
${\bf S}_l(t) \cdot {\bf e}_{\rm y}$.
The left hand side sequence of figures constructed
for $22^{\rm nd}$~stright line 
(solid line) and for $37^{\rm th}$~line 
(dashed line) in the external field direction~${\bf e}_{\rm x}$. 
The right hand side sequence belongs
to $97^{\rm th}$ (solid line) and
$102^{\rm nd}$~line (dashed line)
of the considered
system $D/J=0.1$, $L_0=20$.
}
\end{center}\end{figure}

\begin{figure}[h]\begin{center}
\epsfig{file=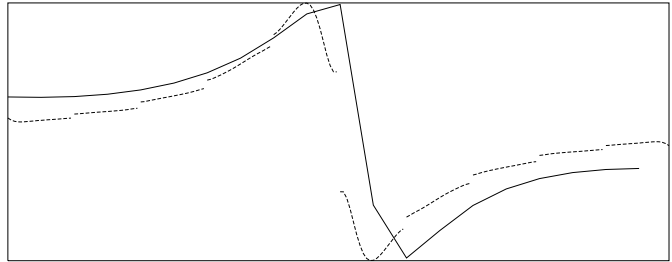,width=9.0cm,angle=0.0}
\caption{\label{sim_soliton}
The snapshot illustrating 
similarity of nonlinear 
WAP
at intra-dot 
($L_0=20$) 
and inter-dot ($L=200$) 
spatial scales 
(at different times).
The rescaling is performed 
for dot indexed 
by $i=55$ 
taken from
$102^{\rm nd}$~line of array 
[spin index changes within $\alpha\in\langle 20500,\,20519\rangle$]
at time $\tau=13.5$ (depicted by solid line)
and for $28^{\rm th}$~spin line of array 
[where $\alpha\in \langle 5600,5799\rangle$]
at $\tau=9.7$ (dashed line). 
The inter-dot and intra-dot
configurations look similar 
when 
the spin 
coordinates are 
transformed using multiplicative 
factor $L_0/L$.
}
\end{center}\end{figure}

\begin{figure}[h]\begin{center}
\epsfig{file=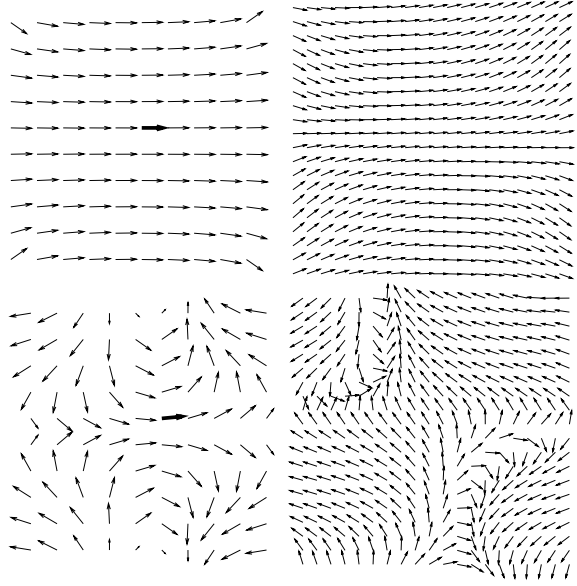,width=8.0cm,angle=0.0}
\caption{\label{sim_dot_array}
The spatio-temporal similarity 
of the magnetic configurations occurring
at intra-dot and array scales.  
On the left hand side we plot
{\em overall moments} of dots 
at different times
$\tau=4.5$ (upper array flower)
and $\tau=8.5$ (bottom array configuration).
The detailed {\em intra-dot} configurations shown
at the right hand side column 
corresponding to dot moments marked
by the bold vector and fixed at the later
times $\tau=8.5$ (upper intra-dot flower)
$\tau=12.5$ (bottom 
three-domain intra-dot configuration)
for $D/J=0.1$, $L_0=20$, $L=200$.}
\end{center}
\end{figure}

At the beginning of reversal
the external magnetic field nucleates reversal domains 
at the corners of dots. The recognition 
of simulated systems shows that remagnetization 
is more homogeneous and easier when dots are localized 
at the corners of array, whereas nonuniformities occur 
due to mirror symmetry of effective fields typical 
for dots standing in positions closer to axial 
zone of array. This property is illustrated 
by snapshot taken at characteristic time 
as is shown in Fig.~\ref{cnf1}. 
The typical for nonuniform 
remagnetization is that reverse 
domains, which gain external field polarity, grow 
from the dot corners 
and pass toward the dot centers. 
At the intermediate 
stages domain  walls - {\em nonlinear waves} 
\cite{Levy1999} 
between reverse and initial polarity 
domains become narrow. In the context 
with nanoparticles and our small-scale 
simulations we prefer to use the term nonlinear wave 
instead of domain wall since nonlinear 
wave better expresses nonlocal character of arising spin 
structures. At the final stage of the inhomogeneous reversal 
the sum of external and dipolar field can, 
if sufficiently strong, homogenize intra-dot ordering 
by annihilating of {\em wave/anti-wave pairs} (WAP). 
Similar mechanism was studied in detail for 
linear spin chain \cite{BraunHB1994}.
As we see from the scheme depicted in Fig.~\ref{model_premag}
the reversal mechanism in dots is position
and time dependent for finite size arrays.
The formation of WAP is an example 
of inhomogeneous remagnetization.
The process is characteristic for later times 
and dots placed nearly the axial array 
zone as is depicted in Fig.~\ref{cnf2}.
The parallelism between 
spin chain and array behavior inspired
the monitoring of spin projections summarized in Fig.~\ref{solitons}. 
The figure shows ${S}_{\alpha y}$ components 
taken from several superlattice site lines crossing array in external 
field direction. The time sequence of these 
projections reveals that each inter-dot interface acts as 
a phase-shifter of WAP formed in different dots. The consequence 
of inter-dot coupling is the large-scale 
WAP with shape resembling its small-scale intra-dot 
counterpart (see Fig.~\ref{sim_soliton}). 
It is rather surprising that 
for its build up magnetostatic forces 
suffice. The spatio-temporal 
{\em two-scale similarity} of magnetic ordering 
illustrated in Fig.~\ref{sim_dot_array} 
originates from spatial 
distribution of dipolar fields 
of square-shaped dots embedded into 
square-shaped array. 

\subsection{The quantitative analysis of switching, 
the effect of bias}

\subsubsection{The dot-size dependences}

\begin{figure}\begin{center}
\epsfig{file=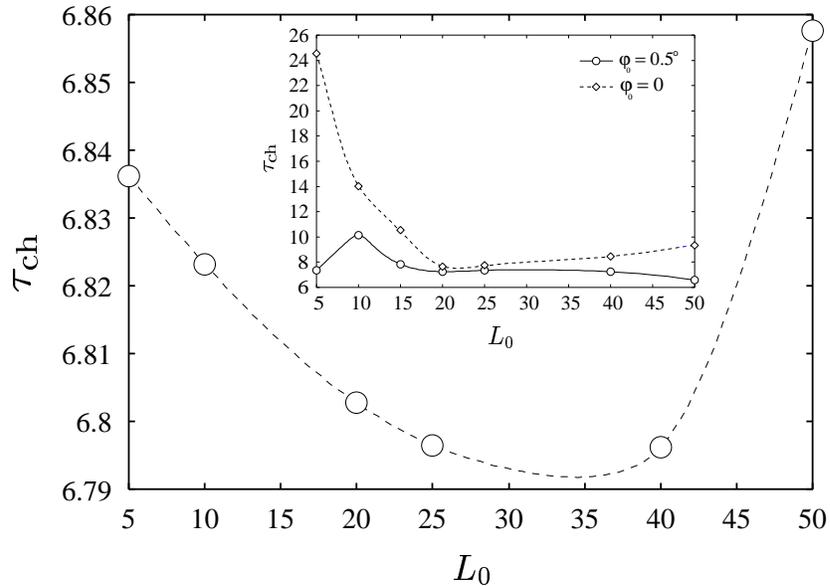,width=11.0cm,angle=0.0}
\caption{\label{t_ch_L_0}
The $\tau_{\rm ch}(L_0)$ 
dependence calculated for 
$D/J=1$, ${\bf H}/J=-{\bf e}_{\rm x}$. 
The insert compares 
characteristic time for $\varphi_0>0$
and $\varphi_0=0$ configurations in 
isolated dot. The most remarkable 
are differences for small $L_0$. 
The lines are guidelines to eye. 
Only a weak dependence 
on hierarchical cut-off 
is observed. 
The test of $\tau_{\rm ch}$ accuracy 
for a given size of array 
is performed for $L_0=5$ 
and  $r_{\rm cut} \in \{\sqrt{2} L_0, 2\sqrt{2} L_0,  3\sqrt{2} L_0\}$.
The calculations confirm that
 changes in $\tau_{\rm ch}$  
 does not exceed 
 bound $0.004$.}
\end{center}\end{figure}

We continue with the comparing of $\tau_{\rm ch}(L_0)$ 
dependences for patterned structure as well as for the isolated dots 
with zero and non-zero bias. The varying of $L_0$ is in some 
sense equivalent to varying of dipolar and 
exchange coupling contributions.  The results for patterning 
$L_0=5, 10, 20, 25, 40, 50$  for $D/J=1$ are depicted in 
Fig.~\ref{t_ch_L_0}, where the characteristic 
times of the interacting and isolated dots 
are also compared.  
From the simulations 
it follows that inhomogeneities 
grow as $L_0$ increases. 
The reverse phase 
nucleates preferentially at dot
surfaces and reduces $\tau_{\rm ch}$ 
in larger dots. 
If inter-dot interactions are considered  
$\tau_{\rm ch}(L_0)$ dependences look smoother than dependences 
obtained for isolated dots.  
The interactions affect spatial dispersion of 
$\omega_{\alpha}$, 
which means that some dots 
are accelerated and some decelerated during reversal.
The simulations of isolated 
dots show that differences 
between zero and non-zero bias are larger for smaller $L_0$. 
The decrease of $\tau_{\rm ch}$ due to uniform bias $\varphi_0>0$ 
is a manifestation of the 
Stoner-Wohlfarth coherent regime.

\subsubsection{  The interplay of 
                 $D$, $J$, $H$, $\varphi_0$ 
                 parameters, 
                 minimal 
                 four-spin model 
              }\label{Interp1}

\begin{figure}\begin{center}
\epsfig{file=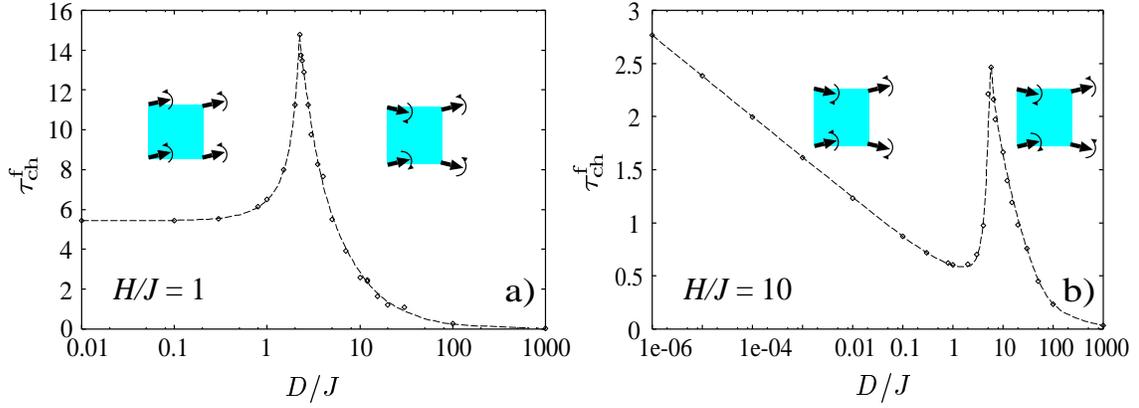,width=15.0cm,angle=0.0}
\caption{\label{fourmers}
The $\tau_{\rm ch}(D/J)$ 
dependences obtained 
for different remagnetization 
(flower and vortex) modes of four-spin 
system simulated for $\varphi_0>0$ 
(left hand side) and for 
$\varphi_0=0$ (right hand side).
The cusp corresponds to four-spin 
bifurcation point 
$D_{\rm c}^{\rm f}$.}
\end{center}
\end{figure}

In this section 
minimal
four-spin model and complete array model 
of magnetization reversal
are discussed and compared. 
The {\em minimal 
model} consists 
of four spins arranged to $2\times 2$ 
square segment of unit size.
In the segment 
four nearest pairs 
are coupled by the exchange interaction 
and all of the pairs by
dipole-dipole interaction. 
Despite of substantial 
geometric truncation, 
all of the interactions 
considered for array
take place in four-spin model.
The simulation of the minimal model 
system 
provides 
$\tau_{\rm ch}^{\rm f}$ 
characteristics depicted 
in ~Fig.~\ref{fourmers}. 
They consist of two 
qualitatively 
different 
regimes 
separated by bifurcation point 
$D=D_{\rm c}^{\rm f}$ 
manifested by 
$\tau_{\rm ch}^{\rm f}(D)$ cusp.  
At $D_{\rm c}^{\rm f}$ several modes compete, 
the interaction and external fields 
are nearly  balanced and  
frustrated spins need 
quite long time to select the winning mode. 
For $D<D_{\rm c}^{\rm f}$ the reversal is very susceptible 
to choice of initial bias. It is easy to check that 
even small $\varphi_0>0$ accelerates 
reversal and stabilizes coherent 
or quasi-coherent rotation 
in coincidence 
with the Stoner-Wohlfarth 
scenario. Further 
simplification of four-spin 
dynamics which clarifies situation
can be done 
assuming the total {\em coherence} of spins.  
The consequence is the elimination 
of three angular 
variables and analytic expression
\begin{equation}
\label{4time}
\tau_{\rm ch}^{\rm f}
=\frac{4\sqrt{2} J}{\cal L}\int_{\varphi_0}^{\pi/2}
\frac{{\rm d}\varphi}
     {3 D\cos(2\varphi)+4\sqrt{2} H\sin\varphi }\,\,\,.
\label{fourmer1}
\end{equation}
where ${\cal L}=
4\pi\lambda/\gamma\mu_0 M_{\rm s}$
is undimensional parameter proportional 
to damping parameter $\lambda$. 
The expression obtained in 
the limit $D\to 0$ is rather well known
\begin{equation}
\label{2time}
\tau_{\rm ch}
=\frac{J}{H \cal L}\int_{\varphi_0}^{\pi/2}
\frac{{\rm d}\varphi}
     {\sin\varphi }= 
\frac{J}{H {\cal L}} 
\ln(\mbox{cotg}(\varphi_0/2))
\,\,\,.
\end{equation}
For $D>0$ the integral Eq.(\ref{fourmer1}) 
leads to finite $\tau_{\rm ch}^{\rm f}$
even in the case $\varphi_0=0$. It means that
dipolar coupling
removes logarithmic
divergence $\propto\ln\varphi_0$ of
$\tau_{\rm ch}^{\rm f}$ 
and substitutes in part the effect of $\varphi_0>0$.
In the regime $D<D_{\rm c}^{\rm f}$
the bias $\varphi_0=0$ induces
non-uniform flower mode 
belonging to characteristic time which 
is out of the scope of Eq.(\ref{fourmer1}).
The four-spin dynamics attained by simulation 
for $D>D_{\rm c}^{\rm f}$
is even more complex. 
In that case the choice of bias 
losses its dynamical relevance. 
The simulations show that 
quite short time
is needed 
to form the transient
long-living flower mode 
(long with respect to
$\tau_{\rm ch}^{\rm f}$) 
which alters 
to vortex mode at final stages.

\begin{figure}\begin{center}
\epsfig{file=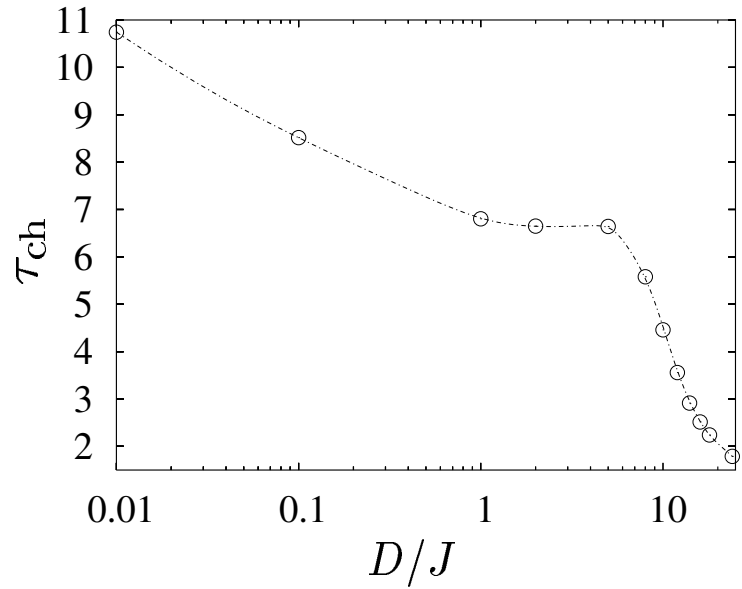,width=11.0cm,angle=0.0}
\caption{\label{tch_vs_DJ}
The plot of 
$\tau_{\rm ch}(D/J)$ 
dependence 
obtained for 
modulation $L_0=20$. 
The  variation of
$\tau_{\rm ch}(D/J)$
slope near $D_{\rm c}$ 
associated with change 
of reversal mode.}
\end{center}\end{figure}

With the four-spin system behavior 
in mind the attention is turned now  
to the {\em complex array dynamics}. 
As we see from Fig.~\ref{tch_vs_DJ}
$\tau_{\rm ch}(D/J)$ dependence does not contain sharp cusp.
Instead of it, 
the slope variation of $\tau_{\rm ch}(D/J)$
arises (labeled as $D_{\rm c}$ 
in analogy to $D_{\rm c}^{\rm f}$).
For $L_0=20$ and $H/J=1$ the
position of $D_{\rm c}$ can be estimated as
$D_{\rm c}\simeq 5 J$. 
The configuration snapshots show 
that near $D_{\rm c}$ 
the array consists of spin 
configuration loops
of diameters approaching single 
dot size or size of several dots.
For $D>D_{\rm c}$ the array reversal
generates non-collinear
antiferromagnetic phases.
The common feature of oversimplified four-spin model 
and array simulation 
is that finite 
$\tau_{\rm ch}^{\rm f}$ and $\tau_{\rm ch}$ 
are attained even for $\varphi_0=0$
due to non-collinearity
of external and dipolar fields.  
From dependences 
depicted 
in Figs.~\ref{t_ch_L_0}~and~\ref{tch_vs_DJ} 
it follows that geometric 
variations in $L_0$ can produce  
less remarkable 
changes than $D/J$ parameters.

\begin{figure}\begin{center}
\epsfig{file=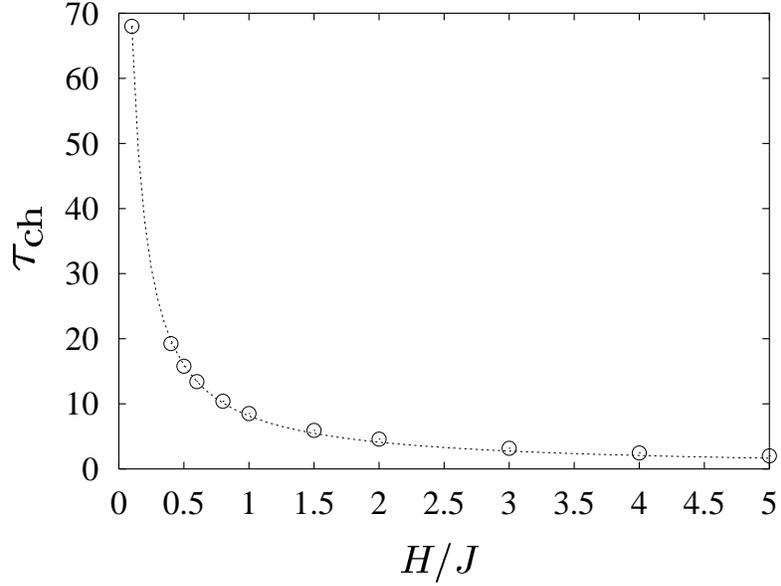,width=11.0cm,angle=0.0}
\caption{\label{tau_ch_HJ}
The plot of $\tau_{\rm ch}(H/J)$ 
dependence 
obtained for 
$L_0=20$ and $D/J=0.1$. 
Dashed line represents 
fit $\tau_{\rm ch}=
a (J/H) - b D\, (J/H^2)$ 
with parameters 
$a=8.302\pm 0.122$ and 
$b=1.506\pm 0.131$. 
The choice 
of fitting function
stems from the asymptotic 
$H$-dependence 
derived for four-spin model 
in sect.\ref{Interp1}.}
\end{center}\end{figure}

For completeness 
the external field dependences 
of characteristic time
should be mentioned.
From Eq.(\ref{fourmer1})
one can obtain for $D/H\ll 1$ 
the asymptotic expansion of characteristic time
$\tau_{\rm ch}^{\rm f}=
(J/({\cal L} H))$ $
\sum_{n=0}^{\infty}
(-1)^n
(D/H)^n
c_n(\varphi_0)
$,
$c_n(\varphi_0)=3^n 2^{-5 n/2}
\int_{\varphi_0}^{\pi/2}{\rm d}\varphi\,
\cos^n(2\varphi)
(\sin( \varphi))^{-n-1}$
suitable for proposition 
of reasonable fitting function 
covering the array data.
The array simulations 
in Fig.~\ref{tau_ch_HJ} shows 
that $\tau_{\rm ch}$
coincides qualitatively 
with asymptotic predictions 
of four-spin system.

\subsection{
The intra-dot 
complex structures in representation of ART network}\label{ART1}

 The artificial neural 
 network models \cite{Haykin98} inspired 
 by the physiology are able to mimic action 
 of neurons  and synaptic connections of brain.  
 The salient feature 
 of neural systems 
 is the associative 
 recognition of the complex structures. 
 This ability is attained by training 
 of synaptic connections - neurons.
 In this paper we deal with networks 
 based on the adaptive resonance theory  (ART) 
 developed by Carpenter and Grossberg {\cite{Carpenter87}}
 originally as a model explaining adaptive phenomena 
 in visual systems. 
 The family of ART algorithms 
 belongs to unsupervised 
 self-organized 
 algorithms motivated 
 by need to construct 
 network sufficiently 
 adaptive to novel inputs.
 The {\em adaptivity} means that 
 network produces new neurons 
 if diversity of inputs 
 overcomes 
 the predefined threshold 
 represented by so 
 called 
 {\it vigilance parameter}. 
 On the other hand 
 redundancy of 
 information leads 
 to the annihilation 
 of neurons. 
 For summary of ART network 
 applications see 
 e.g.~\cite{Honavar94}

\subsubsection{The encoding of intra-dot magnetic structure}

\begin{figure}\begin{center}
 \epsfig{file=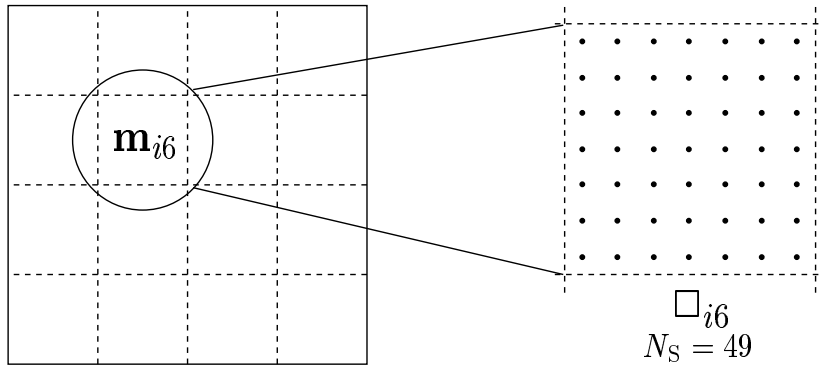,width=11.0cm,angle=0.0}
\caption{\label{spinblok}
 The schematic example of intra-dot 
 segmentation. Figure showing 
 sixth square segment labeled 
 $\Box_{i6}$ belonging to $i$th dot
 including 
 $N_{\rm S}=49$ spins.}
 \end{center}\end{figure}

The ideas of clustering 
and adaptivity of the ART networks 
led us to their application 
to magnetic intra-dot 
configurations 
generated 
during reversal.
At first,
an appropriate format 
for encoding of magnetic configurations is suggested. 
Similarly to formulation 
{\cite{Horvath2001}} 
the configurations 
are encoded by $2\,N_{\rm{c}}$ 
dimensional tuples 
\begin{equation}
  \tilde{m}_i \equiv
 \left[\, {\bf{m}}_{i 1}, 
   {\bf{m}}_{i 2}, \ldots,
   {\bf{m}}_{i N_{\rm{c}}} \right] \,,
\end{equation}
where
\begin{equation}
  {\bf{m}}_{i n} =
\frac{1}{N_{\rm{S}}} \sum_{l \in \Box_{i n}}
  {\bf{S}}_l\,, \quad i = 1, 2, \ldots,
  L^2_{\rm{d}}, \quad n = 1, 2, \ldots, N_{\rm{c}}\,
\label{Eqmin}
\end{equation}
is the local average  
taken from $N_{\rm{S}}$ spins 
(see  Fig.~\ref{spinblok}), 
where sum runs over site $l$ 
taken from $n$th segment 
of $i$th dot labeled as 
$\Box_{i n}$. Each dot is subdivided 
to $N_{\rm{c}}=L_0^2/N_{\rm S}$ 
intra-dot square segments. The system of 
$L_{\rm d}^2$ intra-dot 
configurations is transformed into
$\tilde{m}_i$ tuples incoming to ART, which
compresses them to 
$N_{\rm{w}}\leq L_{\rm d}^2$ neurons 
\begin{equation}
  \tilde{w}_j \equiv [
   {\bf{w}}_{j 1}, {\bf{w}}_{j 2},
   \ldots, {\bf{w}}_{j N_{\rm{c}}} ]
   \,, \qquad j
   \in \{ 1, 2, \ldots N_{\rm w} \}
\end{equation}
representing the basal types of intra-dot ordering. 
The self-organization mechanism of neurons 
discussed in the next 
is  controlled by the coincidence between input 
$\tilde{m}_i$ and output $\tilde{w}_j$. 
As a proper measure of coincidence is chosen Euclidean 
distance
\begin{equation}
  \|{\tilde{w}_j} - \tilde{m}_i \| =
\sqrt{\sum_{n = 1}^{N_{\rm{c}}} (
  {\bf{w}}_{j n} - {\bf{m}}_{i n} ) \cdot ( {\bf{w}}_{j n} -
  {\bf{m}}_{i n} )^{\rm{T}}} \,,
\label{Eucdis}
\end{equation}
where $\rm{T}$ labels transposition.

\subsubsection{The ART clustering algorithm}

The training algorithm is described
in the following points
\begin{enumerate}
  \item[{\bf 1.}] {\it Initialization} 
 setting $N_{\rm{w}} = 1$, $\tilde{w}_1
  = \tilde{m}_i$ for 
 random dot selection between $i \in \{ 0, 1,\ldots,  L_{\rm{d}}^2-1 \}$.

  \item[{\bf 2.}] {\it Loop} through 
  the set of intra-dot
  magnetic configurations. For each randomly selected dot $i$ follows:
  \begin{enumerate}
    \item[{\bf 2.1}] {\it Presentation} of $\tilde{m}_i$
    to ART network. The training is realized
    for $k = 1, 2, \ldots$ iteration steps.

    \item[{\bf 2.2}] {\it Computing}
   of the actual
   index 
   $j^{\ast} ( k, i )$ of the 
   {\it fired neuron}
   $\tilde{w}_{j^{\ast}(k,i)}(k)$ 
   trained by 
   ${\tilde m}_i$
   pattern according 
   to the competitive 
   rule
    \begin{equation}
      \begin{array}{rlcl}
       j^{\ast} ( k, i ) = & \mbox{arg}&\mbox{min} &  {\| {\tilde{w}_j} ( k ) -\tilde{m}_i \|} \\
       & & ^{j = 1,2,\ldots, N_{\rm{w}}} &
     \end{array}\,\,\,.
    \end{equation}

    \item[{\bf 2.3}] {\it Comparing} $\|{\tilde{w}_{
      j^{\ast}(k,i)}} ( k ) -
      \tilde{m}_i \|$ to the
      vigilance parameter $\rho$.
      \begin{enumerate}
      \item[{\bf 2.3.1}] {\it Update} 
       of the neuron weights via Hebbian 
       training rule \cite{Haykin98}
      \begin{equation}
       \tilde{w}_{j^{\ast}(k,i)} ( k + 1 ) =
       \tilde{w}_{j^{\ast}(k,i)} ( k ) +
       \eta ( k ) \,
      \left( \tilde{m}_i - \tilde{w}_{j^{\ast}(k,i)} ( k ) \right)
      \end{equation}
      is applied if
      $\|{\tilde{w}_{j^{\ast}(k,i)}} - \tilde{m}_i \| \leq \rho$.
      The update shifts winning 
      vector toward nearby input
      with rate proportional 
      to plasticity parameter 
      $ \eta(k) = \eta_0 \exp 
      \left(- k/\tau_{\rm{train}} \right)$ 
      relaxing with 
      training time 
      $\tau_{\rm train}$.

      \item[{\bf 2.3.2}] {\it Creation} 
      of new neuron
      $N_{\rm{w}} \leftarrow N_{\rm{w}} + 1$,
      $\tilde{w}_{_{N_{\rm{w}}}} \leftarrow \tilde{m}_i$
      if 
      $\|{\tilde{w}_{j^{\ast}(k,i)}}(k) - \tilde{m}_i \| > \rho$.

    \end{enumerate}
    \item[{\bf 2.4}] {\it Annihilation}
     of neuron pair $[\,{\tilde w}_{z_1^{\ast}}(k),
     {\tilde w}_{z_2^{\ast}}(k)\,]$,
     ordered as $z_1^{\ast}< z_2^{\ast}$,
     with indices $z_1^{\ast}$, $z_2^{\ast}$
     selected according to minimum distance
    \begin{equation}
     \begin{array}{rlcl}
     [ z_1^{\ast}, z_2^{\ast} ] =
     & \mbox{arg}
     & \mbox{min}
     & {\|{\tilde{w}_{z_1}}-\tilde{w}_{z_2} \|}
     \\
     & & ^{z_1, z_2} &
     \end{array}
    \end{equation}
    if $\|{\tilde{w}_{
    z_1^{\ast}}}-\tilde{w}_{z_2^{\ast}} \|<\rho\,$.
    The product of annihilation 
    $[ z_1^{\ast}, z_2^{\ast} ] \rightarrow z_1^{\ast}$,
    $N_{\rm{w}} \leftarrow  N_{\rm{w}} - 1$
    is the neuron
    $\tilde{w}_{z_1^{\ast}}$
    determined by the midpoint rule
    $\tilde{w}_{z_1^{\ast}}
    \leftarrow \frac{1}{2} ( 
     \tilde{w}_{z_1^{\ast}} +
    \tilde{w}_{z_2^{\ast}} )$.
    After this update of
    $\tilde{w}_{z_1^{\ast}}$ 
    the neurons having 
    $j\geq z_2^{\ast}$  
    are rewritten: 
    ${\tilde w}_{j-1}\leftarrow {\tilde w}_{j}$.

  \end{enumerate}
  \item[{\bf 3.}] {\it Stop criterion} 
 is represented
  by the inequality
  \begin{equation}
    \frac{1}{L_{\rm{d}}^2}
   \sum_{i = 0}^{L_{\rm{d}}^2-1}
    \|
    {\tilde{w}_{j^{\ast}( k , i )}}(k+1) -
    \tilde{w}_{j^{\ast} ( k , i )}(k) \| <
    \varepsilon\,\,,
  \end{equation}
  where $\varepsilon$ is small parameter.
  If the above inequality is not fulfilled, the algorithm
  follows from the step {\bf 2} with $k$ incremented by $1$.
  \end{enumerate}
The stop criterion 
indicates when the network attains 
a fixed point of 
${\tilde w}_j,\,j=1,2, \ldots, N_{\rm w}$ including the cluster 
of typical representative intra-dot configurations.

\subsubsection{The ART applied 
               to different stages 
               of reversal}

\begin{figure}\begin{center}
\epsfig{file=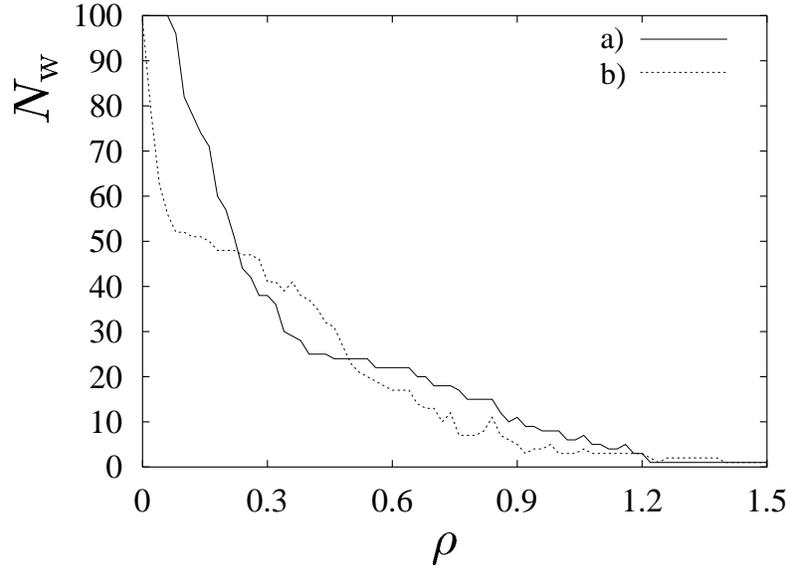,width=11.0cm,angle=0.0}
\caption{\label{vigNN}
The pruning of ART neurons. 
The number of neurons $N_{\rm w}$ 
decreases with increasing vigilance parameter. 
The dependences obtained for 
$D/J=0.1$ and $L_0=20$. 
The intra-dot configuration is locally averaged over 
$N_{\rm S}=4$ spins [see Eq.(\ref{Eqmin})].
Case~a) corresponds to configuration near 
to the switching time 
($\tau= 8.5$).
Case~b)~belongs to time when the internal 
energy (free from Zeeman term) is near its maximum 
($\tau=11$).
}
\end{center}
\end{figure}

\begin{figure}\begin{center}
\epsfig{file=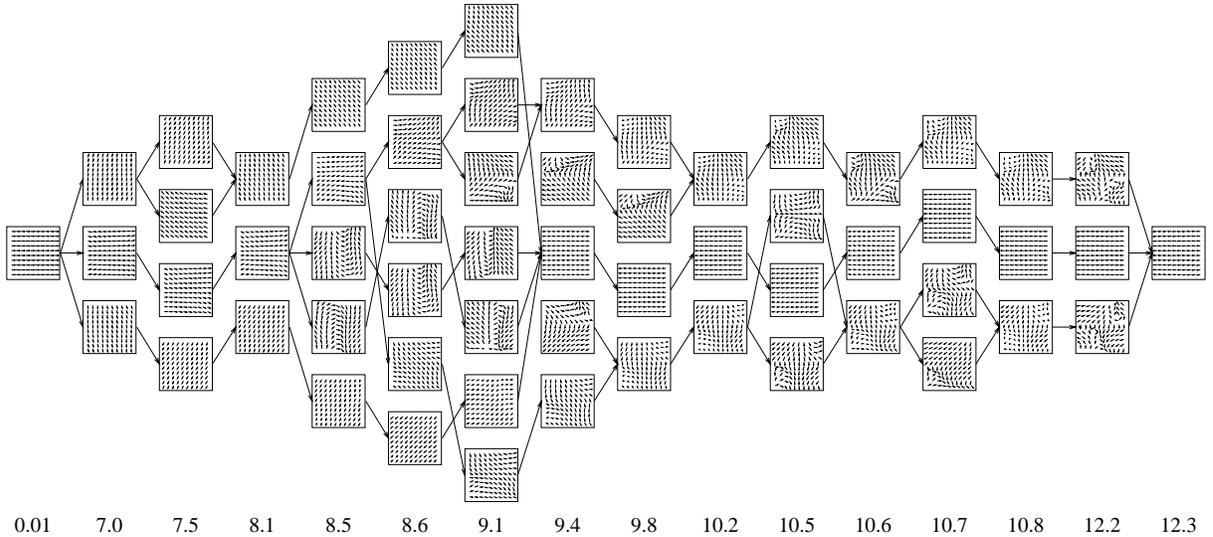,width=16.0cm,angle=0.0}
\caption{\label{diagartNN}
The ART network representation 
of the main
intra-dot transients of array 
reversal 
for   
$D/J=0.1$, $L_0=20$.
The configurations 
corresponding to different times 
obtained for 
$N_{\rm S}=4$ segment 
average [see Eq.(\ref{Eqmin})]. 
The arrows 
connect closest
transient configurations 
(in the sense of 
the magnetic Euclidean distance).
The extremal number of neurons
occurring for 
$\tau=9.1$ 
roughly corresponds 
to the maximum 
of internal energy.}
\end{center}
\end{figure}

\begin{figure}\begin{center}
\epsfig{file=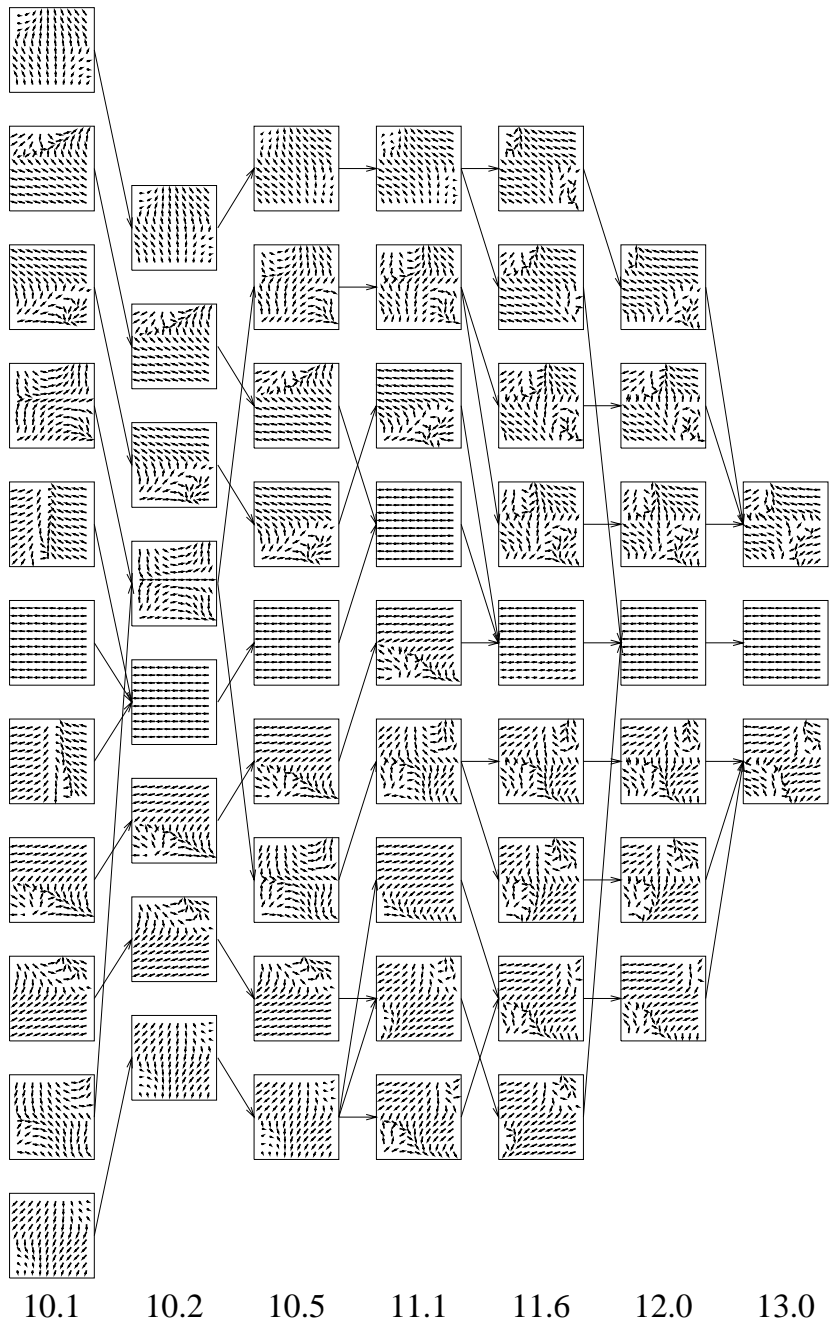,width=11.0cm,angle=0.0}
\caption{\label{detailNN}
Figure showing ART 
trained with relatively small 
vigilance $\rho=0.8$.  
The network properly 
uncovers mechanism of WAP 
annihilation and nonlinear wave motion.}
\end{center}
\end{figure}

\begin{figure}\begin{center}
\epsfig{file=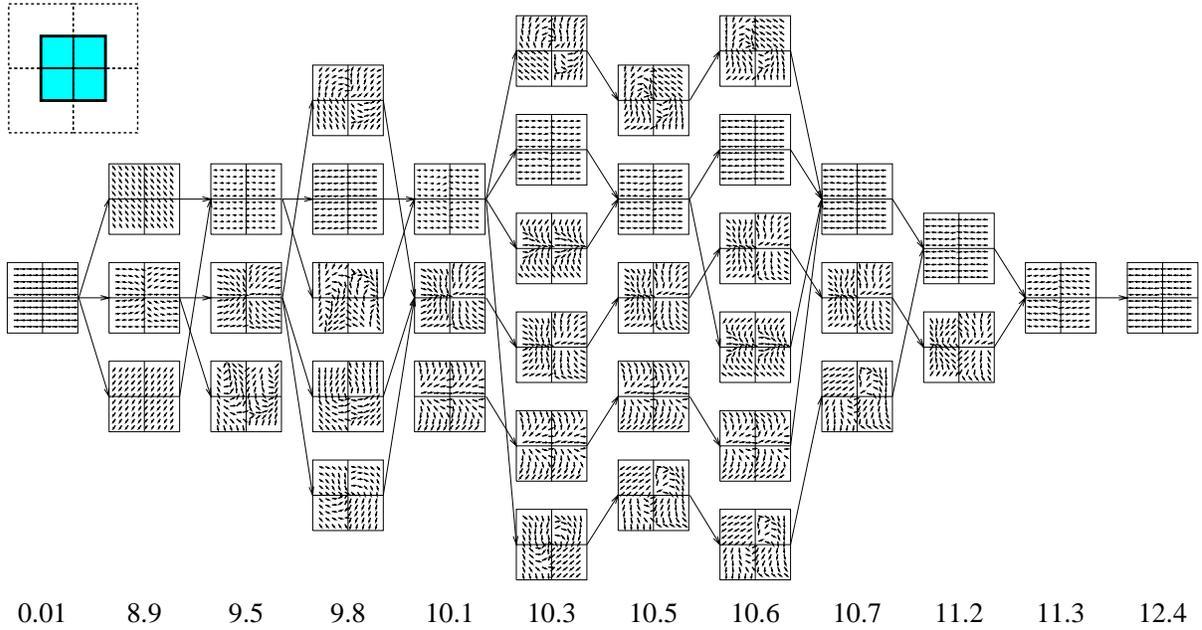,width=16.0cm,angle=0.0}
\caption{\label{interface}
 The ART revealing 
 interfacial 
 four-dot modes of  
 magnetization reversal: 
 inter-dot nonlinear 
 wave pinning and formation of 
 inter-dot vortices.
 Parameters:
 $D/J=0.1$,  
 $\rho=1.22$, 
 $N_{\rm S}=4$, 
 $L_0=10$. 
} 
\end{center}
\end{figure}

The ART network has been applied
independently to array configurations
attained at selected stages of reversal.
The results of self-organization are presented 
in Figs.~\ref{diagartNN}-\ref{interface}.
The ART self-organization 
was carried out for tuned parameters
$\eta_0=0.1$, $\tau_{\rm train}=20$,
$\varepsilon=10^{-6}$, $\rho=1.15$. 
The most important seems to be 
optimized choice of an appropriate vigilance connecting 
$\rho$ and $N_{\rm w}$. 
It stem from calibration in Fig.~\ref{vigNN}. 
During it $\rho$ was decreased slowly 
from the initial $\rho=1.5$ stabilizing single neuron. 
This decrease of $\rho$ 
was stopped for 
$N_{\rm w}=L_{\rm d}^2=100$ competing neurons.

The ART view point on reversal leads 
to intra-dot taxonomy depicted in Fig.~\ref{diagartNN}. 
In this figure the most similar neurons 
[in the sense of generalized distance Eq.(\ref{Eucdis})]
at different times are joined by the arrows. 
The ART model clearly 
distinguishes between branches of 
coherently rotating monodomain 
dots and dots with more complex order. 
The  diversity of configurations occurring near the internal energy 
maxima is reflected by the cardinality of neuron population. 
The ART pruning at late time stages 
is associated with small intra-dot 
diversity within the array. 
How choice of $\rho$ affects the information 
content of network is clearly demonstrated in Fig.~\ref{detailNN}. 
The supplementary problem solved by ART represents 
classification of interfacial inter-dot configurations.  
In that case the configuration tuples 
${\tilde m_i}$,  
${\tilde w}_j$ 
have been combined from ${\bf m}_{in}$ 
segments 
[see Eq.(\ref{Eqmin})] 
of adjacent quadrants 
of four neighboring dots. 
The results depicted in Fig.~\ref{interface} 
indicate preferential formation 
of inter-dot vortices and strong 
tendency to inter-dot pinning
also for conditions when formation 
of intra-dot vortices is disadvantageous. 

\section{Conclusions}

The magnetization 
reversal has been simulated for
superlattice spin model of the ultra-dense 
magnetic dot array.  
Several aspects 
of reversal dynamics for different 
exchange and magnetostatic couplings,  
external fields, geometric parameters 
and initial conditions 
have been studied.
It was found that the principal 
physical consequence of geometric modulation 
is the formation of two-scale WAP.  
The ART viewpoint 
seems to be fruitful in thinking 
how dot switches from one configuration 
to another within 
the large time intervals, 
and how to identify 
dominant channels of  
intra-dot evolution. 
The present experience creates 
believe that advantages 
of ART paradigm should 
be exploited namely in 3-D lattices, 
where visual classification 
meets natural bounds.

\section{Acknowledgments}
The work was partially 
supported by the 
grant No.1/6020/99 (Slovak Grant Agency VEGA). M.~Gmitra expresses 
thanks to Dr.~Z.~Frait and Dr.~V.~Kambersk\'y from Institute of Physics 
AS CR, Prague, for their hospitality 
and valuable discussions during stay 
supported by training network contract 
HPRN-CT-1999-00150.

\appendix
\section{Appendix}
\subsection{One dimensional enumeration scheme 
            of superlattice 
            sites}\label{1d_enumeration}

\begin{figure}[h]
\begin{center}
\epsfig{file=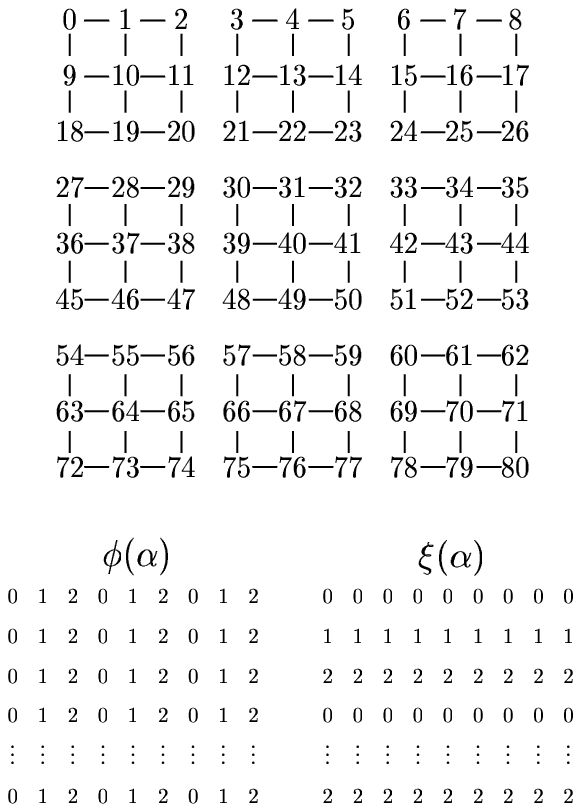,width=7.0cm,angle=0.0}
\caption{\label{modfunphipsi}
The illustration 
    of one-dimensional
    site enumeration 
    scheme 
    ($L_0=3$, $L=9$).
      The modulation
      of connectivity matrix 
      $\epsilon_{\alpha\beta}$ 
      with period $L_0=3$
      in  
      ${\bf e}_{\rm x}$ and 
      ${\bf e}_{\rm y}$
      perpendicular 
      directions 
      realized
      by 
      functions 
      $\phi(\alpha)$, 
      $\xi(\alpha)$ 
      from
      Eq.(\ref{Eqxi}).}
\end{center}
\end{figure}

In the Appendix site enumeration scheme mentioned in 
section~\ref{xy_model} is described. 
In this scheme the superlattice 
relative position vector 
${\bf r}_{\alpha\beta}$
pointing from site 
$\alpha$ to site 
$\beta$ is equal to
\begin{eqnarray}
{\bf r}_{\alpha\beta}
=\left[\,{\cal M}(\beta,L)-{\cal M}(\alpha,L)\,\right]{\bf e}_{\rm x} +
                       \left[\,{\cal I}(\beta,L)-{\cal I}(\alpha,L)\,\right]{\bf e}_{\rm y}
\,,
\end{eqnarray}
where 
\begin{equation}
{\cal M}(a,b) = 
a\, \mbox{mod}\, b \,,\qquad
{\cal I}(a,b) =
\left[a/b\right]_{\mbox{\tiny int}}\,.
\end{equation}
Here ${\cal M}$ is standard modulo function 
and ${\cal I}$ is the integer part of its argument
($L$ is an integer multiple of $L_0$).  
The second part of 
Eq.(\ref{ExchEq}) includes 
the exchange coupling 
$J\epsilon_{\alpha\beta}$ generated by dependence
\begin{equation}
\epsilon_{\alpha\beta}
=
\left(\,
1-\delta_{\phi(\beta)-1,\phi(\alpha)-L_0}
\right)
\left(\,
1-\delta_{\xi(\beta)-1,\xi(\alpha)-L_0}\,
\right)\,.
\label{Eqxi}
\end{equation}
It gives rise to geometry of dense square patterning  
of the period $L_0$ 
in perpendicular 
${\bf e}_{\rm x}$, 
${\bf e}_{\rm y}$ directions. 
In Eq.(\ref{Eqxi}) $\delta$ 
is the Kronecker symbol and 
\begin{equation}
\phi(\alpha) = {\cal M}({\cal M}(\alpha,L),L_0)\,,
\qquad
\xi(\alpha) = 
{\cal M}({\cal I}(\alpha,L),L_0)\,
\label{Eqxi1}
\end{equation}
are dependences 
illustrated in Fig.~\ref{modfunphipsi}. 
The position 
of some dot 
including site 
$\alpha$ can be also 
written in terms of ${\cal M}$ 
and ${\cal I}$ functions:
\begin{equation}
{\bf r}^{\rm d}_{i(\alpha)}=
{\cal I}({\cal M}(\alpha,L),L_0){\bf e}_{\rm x}
+{\cal I}({\cal I}(\alpha,L),L_0) {\bf e}_{\rm y}\,.
\end{equation}
Here
\begin{equation}
i(\alpha)=L_{\rm d}{\cal I}({\cal I}(\alpha,L),L_0)+
{\cal I}({\cal M}(\alpha,L),L_0)\,
\end{equation}
is one-dimensional 
map associating site 
index 
$\alpha$ with dot 
index 
$i(\alpha)\in \{0,1,2,\ldots,L_{\rm d}^2-1\}$.

\subsection{Integration scheme}
\label{integ_scheme}
To derive explicit integration scheme 
with variable
time steps 
$\kappa_1 \Delta \tau$ and 
$\kappa_2 \Delta \tau$
it is supposed 
that integral 
of spin angular frequency
${\overline \omega}_{\alpha}(\tau)$ 
is approximated by
\begin{equation}
\int_{\tau}^{\tau+\Delta \tau}
{\omega}_\alpha(\tau')\,
\mbox{d}\tau'
=
\Delta \tau\,{\overline \omega}_{\alpha}(\tau)\,\,\,,
\label{Eqangu1}
\end{equation}
where
\begin{equation}
\overline {\omega}_{\alpha}(\tau)=
a_0\,\omega_\alpha(\tau)
+a_1\,\omega_\alpha(\tau-\kappa_1 \Delta \tau)
+a_2\,\omega_\alpha(\tau-\kappa_1 \Delta \tau-\kappa_2\Delta \tau)
\end{equation}
is $\tau-$local $\Delta \tau$ 
averaged $\omega_{\alpha}(\tau')$ expressed 
through coefficients
\begin{eqnarray}
a_0 &=& \frac{  2+ 6 \kappa_1^2+ 3 \kappa_2+ 6 \kappa_1 (1+\kappa_2)}{ 6
\kappa_1 (\kappa_1+\kappa_2)}\,, \\
a_1 &=& -\frac{ 2 + 3 \kappa_1+ 3 \kappa_2}{
6 \kappa_1 \kappa_2}\,, \\
a_2 &=& \frac{  2+ 3 \kappa_1}{6 \kappa_2 (\kappa_1+\kappa_2)}\,
\end{eqnarray}
derived by equating to zero the series of difference
$[{\rm lhs.Eq.}(\ref{Eqangu1}) -
  {\rm rhs.Eq.}(\ref{Eqangu1})]$
considered up to the order
${\cal O}(\Delta \tau^4)$.
The reason for use 
of non-constant 
prefactors 
$\kappa_{1,2}\sim 1$ is the 
opportunity to change adaptively 
integration step via the 
instantaneous spin rotation extreme
\begin{equation}
\begin{array}{cl}
{\rm max} & \overline{\omega}_{\alpha}(\tau-\kappa_1 \Delta \tau)\,.\\
{^{\alpha=0,\ldots,L^2-1}} &
 \end{array}
\end{equation}
 In the uniform case $\kappa_1=\kappa_2=1$, 
 the quadrature is compatible 
 with Adams-Bashworth 
 3th order integration formula. 
 The initial two 
 steps 
 estimating $\omega_{\alpha}(\tau-\Delta \tau)$ and
 $\omega (\tau - 2\Delta \tau)$
 can be performed 
 using Euler scheme obtained for $a_0=1$, 
 $a_1=a_2=0$.


\newpage

\section{Figure Captions}
\vspace{1cm}





\end{document}